\documentclass[pp]{aastex}

\begin{document}

\def\Res   {${\cal R}$}
\def\FR	   {${\cal F}$}
\def\Fin   {${\cal N}$}
\def\FinE  {${\cal N}_E$}
\def\FinA  {${\cal N}_A$}
\def\FinD  {${\cal N}_D$}
\def\FinR  {${\cal N}_R$}
\def\B     {${\cal B}$}
\def\Ha    {H$\alpha$}
\def\Hb    {H$\beta$}
\def\NIIb  {[N${\scriptstyle\rm II}$]$\lambda$6548}
\def\NIIr  {[N${\scriptstyle\rm II}$]$\lambda$6583}
\def\OIIs  {[O${\scriptstyle\rm II}$]$\lambda$3727}
\def\OIIIr {[O${\scriptstyle\rm III}$]$\lambda$5007}
\def\SIIb  {[S${\scriptstyle\rm II}$]$\lambda$6717}
\def\NII   {[N${\scriptstyle\rm II}$]}
\def\OI    {[O${\scriptstyle\rm I}$]}
\def\OII   {[O${\scriptstyle\rm II}$]}
\def\OIII  {[O${\scriptstyle\rm III}$]}
\def\SII   {[S${\scriptstyle\rm II}$]}
\def\SiII  {Si${\scriptstyle\rm II}$}
\def\HeI   {He${\scriptstyle\rm I}$}
\def\HII   {H${\scriptstyle\rm II}$}
\def\HI    {H${\scriptstyle\rm I}$}
\def\CaII  {Ca${\scriptstyle\rm II}$}
\def\MgII  {Mg${\scriptstyle\rm II}$}
\def\CII   {C${\scriptstyle\rm II}$}
\def\CIV   {C${\scriptstyle\rm IV}$}

\def\Hzero  {H$^{\scriptscriptstyle o}$}
\def\Nzero  {N$^{\scriptscriptstyle o}$}
\def\Ozero  {O$^{\scriptscriptstyle o}$}
\def\Hplus  {H$^{\scriptscriptstyle +}$}
\def\Nplus  {N$^{\scriptscriptstyle +}$}
\def\Oplus  {O$^{\scriptscriptstyle +}$}
\def\Hpplus {H$^{\scriptscriptstyle ++}$}
\def\Npplus {N$^{\scriptscriptstyle ++}$}
\def\Opplus {O$^{\scriptscriptstyle ++}$}

\def\deg {$^\circ$}
\def\arcmin {$^\prime$}
\def\arcsec {$^{\prime\prime}$}
\def\spose#1{\hbox to 0pt{#1\hss}}
\def\simlt{\mathrel{\spose{\lower 3pt\hbox{$\mathchar"218$}}
    \raise 2.0pt\hbox{$\mathchar"13C$}}}
\def\simgt{\mathrel{\spose{\lower 3pt\hbox{$\mathchar"218$}}
    \raise 2.0pt\hbox{$\mathchar"13E$}}}
\def\Em{{${\cal E}_m$}}
\def\emunits{{\,cm$^{-6}$\,pc}}
\def\eg{{\rm e.g., }}
\def\qv{{\rm q.v., }}
\def\aB{{\alpha_{\rm B}}}
\def\Pcos{{$\Phi^\circ$}}
\def\Jcos{{$J_{-21}^\circ$}}
\def\phiI{{$\phi_i$}}
\def\ie{{\rm i.e. }}
\def\etal{{\rm\ et~al. }}
\def\tauLL{{$\tau_{\scriptscriptstyle LL}$}}
\def\Vlsr{{${\rm V_{LSR}}$}}
\def\Vgsr{{${\rm V_{GSR}}$}}
\def\kms{{\,km\,s$^{-1}$}}
\def\Msun{{\,M$_\odot$}}

\def\aj{{AJ}}
\def\araa{{ARA\&A}}
\def\apj{{ApJ}}
\def\apjs{{ApJS}}
\def\apss{{Ap\&SS}}
\def\aap{{A\&A}}
\def\aapr{{A\&A~Rev.}}
\def\aaps{{A\&AS}}
\def\azh{{AZh}}
\def\jrasc{{JRASC}}
\def\mnras{{MNRAS}}
\def\pasa{{PASA}}
\def\pasp{{PASP}}
\def\pasj{{PASJ}}
\def\sovast{{Soviet~Ast.}}
\def\ssr{{Space~Sci.~Rev.}}
\def\zap{{ZAp}}
\def\nat{{Nature}}
\def\aplett{{Astrophys.~Lett.}}
\def\fcp{{Fund.~Cosmic~Phys.}}
\def\memsai{{Mem.~Soc.~Astron.~Italiana}}
\def\nphysa{{Nucl.~Phys.~A}}
\def\physrep{{Phys.~Rep.}}

\newcommand{\deltam}{\langle \delta \rangle}
\def\la{\mathrel
{\hbox{\rlap{\hbox{\lower3pt\hbox{$\sim$}}}\raise2pt\hbox{$<$}}}}
\def\ga{\mathrel
{\hbox{\rlap{\hbox{\lower3pt\hbox{$\sim$}}}\raise2pt\hbox{$>$}}}}
\newcommand{\dd}{\: {\rm d}}
\newcommand{\half}{\frac{1}{2}}

\title{A Tunable Lyot Filter at Prime Focus: a Method for Tracing Supercluster Scales at $z\sim 1$}

\author{J. Bland-Hawthorn}
\affil{Anglo-Australian Observatory, P.O. Box 296, Epping, NSW 2121, Australia}
\author{W. van Breugel}
\affil{Institute of Geophysics \& Planetary Physics, Lawrence Livermore National Labs, Livermore, USA}
\author{P.R. Gillingham, I.K. Baldry}
\affil{Anglo-Australian Observatory, P.O. Box 296, Epping, NSW 2121, Australia}
\author{D.H. Jones}
\affil{European Southern Observatory, Casilla 19001, Santiago 19, Chile}

\begin{abstract}
Tunable narrow-band, emission-line surveys have begun to show the ease with which star
forming galaxies can be identified in restricted redshift intervals to $z\sim 5$
with a 4m class telescope.  These surveys have been carried out with imaging systems
at the Cassegrain or Nasmyth focus and are therefore restricted to fields smaller than 
10\arcmin.  We now show that tunable narrowband imaging is possible over a 30\arcmin\ field
with a high-performance Lyot filter placed directly in front of a CCD
mosaic at the prime focus.  Our design is intended for the f/3.3 prime focus of the AAT~3.9m, 
although similar devices can be envisaged for the Subaru~8m (f/2), Palomar~5m (f/3.4),
VISTA~4m (f/6), Mayall~4m (f/2.6) or CFHT~3.6m (f/4). 
A modified Wynne doublet ensures sub-arcsecond performance over the field. In 
combination with the new Wide-Field Imaging 8K$\times$8K mosaic (WFI) at the AAT, the overall 
throughput (35\%) of the system to unpolarised light is expected to be comparable to the
TAURUS Tunable Filter (TTF).  Unlike the TTF, the field is fully monochromatic and the
instrumental profile has much better wing suppression.  For targetted surveys of emission-line 
sources at $z \sim 1$, a low-resolution (${\cal R} \sim 150$ at 550nm) Lyot filter on 
a 4m telescope is expected to be comparable or superior to current instruments on 8$-$10m 
class telescopes.  We demonstrate that the 30\arcmin\ field is well matched to superclusters 
at these redshifts such that large-scale structure should be directly observable.  
\end{abstract}


\section{Introduction}
Wide-field imaging surveys continue to dominate many aspects of modern astrophysics.
This holds true for searches of rare objects (\eg brown dwarfs, quasars) or 
high-density sources (\eg star-forming objects at cosmological redshift). Traditional
all-sky surveys in broad bands have been effective in identifying several classes 
of astrophysical sources (\eg photographic surveys with Schmidt telescopes).  The 
Sloan Digital Sky Survey (Gunn\etal\ 1998) promises to identify a wider class of sources
through the use of drift-scanning with a multi-band camera mosaic, which results in 
much smaller differential errors between photometric bands.

In a new development, cosmological surveys of emission-line sources (\eg Hu, Cowie \& McMahon 1998; 
Stockton\etal\ 1999; Kudritzki\etal\ 2000; Jones \& Bland-Hawthorn 2001; Baker\etal\ 2001) underline 
the great potential of targetting narrow photometric bands, particularly those which are directed 
at fields with pre-existing, broadband data.  Indeed, tunable filter devices are under
development for the ESO/VLT, Gemini and GranTeCan telescopes with the primary aim of
finding the highest redshift galaxies.  To date, the maximum field of view available 
to existing or planned instruments is less than 10\arcmin.

We anticipate that future wide-field surveys at $z\sim 1$ will need to reach degree 
scales in order to trace the evolution of large-scale structure.  There are numerous 
arguments which support this. In a low density universe, density fluctuations cease to grow 
below $1+z \sim (\Omega_o^{-1} - 1)^\rho$ ($\rho=1$, $\Omega_o < 1$; $\rho={1\over 3}$, 
$\Omega_o=1$, $\Omega_{\Lambda} > 0$) 
after which time the evolution is very slow (Peacock 1999). For an $\Omega_o = 1$ universe, 
clusters continue to evolve to the present day.  If one considers galaxies on turn-around 
orbits which are just beginning to separate from the general Hubble flow, 
the sphere of influence of a Virgo-like potential projects to $0.5 a/(10 h^{-1} {\rm\ Mpc})$~deg 
at $z\sim 1$ ($\Omega_m=0.3$, $\Omega_\Lambda = 0.7$, $h = H_o/(100\ {\rm km s^{-1}\ Mpc^{-1}})$), 
where $a$ is the transverse radius in comoving coordinates (\eg Kaiser 1987; Eke, Cole \& Frenk 1996).
If we consider the size of the sphere in an unperturbed region of the universe
which encompasses the mass of a rich cluster ($\sim 5\times 10^{14} h^{-1} $M$_\odot$),
then $a\: \sim\: 8\: h^{-1}$ Mpc.  This is consistent with N-body simulations by the GIF and VIRGO 
consortia which reveal that supergalactic structure is anticipated on scales of 0.5\deg\ or more
(Kauffmann\etal\ 1999; Jenkins\etal\ 1998\footnote{See
http://star-www.dur.ac.uk/$\sim$ frazerp/virgo/virgo.html}).
If star-forming, dwarf galaxies or Population III star clusters trace the 
`foothills' of large-scale structure (Benson\etal\ 2000), it is not unreasonable to expect that 
future narrowband surveys on degree scales could directly observe the supercluster networks
(cf. Palunas, Francis \& Woodgate 2000).
But in order to achieve a representative sample, galaxy formation `bias' argues for an even larger 
survey scale. If we consider that growing bias at high redshift partly compensates for evolution in 
the spectrum of mass fluctuations (Peacock 1999), we need to reach a minimum co-moving dimension of 
$\sim100\: h^{-1}$ Mpc which subtends about three degrees! This is far beyond the reach of modern
survey spectrographs which have fields of view an order of magnitude smaller.

The ``area-solid angle'' (${\cal A}.\Omega$) product is a useful measure of survey 
efficiency.  This quantity depends inversely on the square of the telescope f/ratio
(see below) which is unfortunate since narrow spectral bands are seriously degraded in fast 
beams.  Prima facie, this appears to argue against tunable filters as efficient cosmological 
surveyors.  However, remarkably, Lyot (1933, 1944) showed that beams as fast as f/2 could be 
compensated by crossed birefringent elements such that even sub-Angstrom, wide-field 
images are possible. This forms the basis of the tunable Lyot filter design presented 
here.

Traditionally, wide-field surveys have been restricted to broad photometric bands. 
Substantially narrower bands provide a technical challenge.  Survey efficiency is 
often loosely stated in terms of the ${\cal A}.\Omega$ product, where ${\cal A}$ is the 
telescope area, and $\Omega$ is the total solid angle of the sky survey. 
For a single telescope pointing, 
\begin{equation}
{\cal A}_1.\Omega_1 \approx {\cal A}_2.\Omega_2 \approx ({\frac{\pi^2 d^2}{4 {\cal F}^2}})
\label{eq:A-Omega}
\end{equation}
where $d$ is the detector size in mm, and \FR\ is the beam focal ratio 
(0.5~$\leq$~\FR~$<$~$\infty$).  The subscripts are defined in Fig.~\ref{fig:A-Omega}.
For ${\cal A}.\Omega$ to be a useful measure of survey efficiency, there are 
certain qualifications. We assume that the sources under study are (a) resolved by
the instrument and the pixel sampling, (b) detected in a reasonable exposure time, and 
(c) the signal-to-noise ratio (SNR) increases as (time)$^{\frac{1}{2}}$.

For any telescope, fast beams are much more efficient for survey work. 
In eqn.~\ref{eq:A-Omega}, to sidestep subtleties involving field-expanded foci, we adopt 
\FR~$\geq$~2 as faster beams cannot be properly compensated by Lyot's method.
But fast beams degrade the energy interval isolated through spectroscopic interference,
particularly for interference filters.
Interestingly, this is not aided by going to collimated beams since the focal
reducer preserves the field angle $\theta$ of the incoming beam. Since the path 
length has an angular dependence, interference filters exhibit a phase effect 
(shifting central wavelength) over the field of view. An impractical solution,
but not without precedent,
is to produce an interference filter which covers the entrance aperture of the 
telescope. The UK Schmidt 35cm \Ha\ (${\cal R} \sim 100$) filter (Parker \&
Bland-Hawthorn 1997) is the largest and most expensive interference filter made 
for astronomical purposes. By a simple scaling, a single monolithic filter for a 
1m class telescope would cost in excess of US\$1M, rather more than the 
cost of the fully tunable optical system (filter $+$ top end) proposed here.

In $\S$2 and $\S$3, we derive the degradation of interference filters and etalon 
filters in a fast beam. In $\S$4, we describe the structure and 
operation of the tunable Lyot filter, and the underlying mechanism of beam 
compensation is summarised. The ray-traced optical path of a wide-field Lyot 
filter is presented in $\S$5, with particular application to the AAT f/3.25 Prime 
Focus. The fully corrected field requires a modified Wynne doublet. In $\S$6, 
we present a brief science case for a monochromatic, tunable imaging filter.

\section{Narrowband filters in a converging beam}

Earlier work on filter degradation in fast beams (\eg Lissberger \& Wilcock 1959)
derive geometric approximations or resort to simulation. In fact, the effect is
fully analytic. In Appendix A, we provide a complete derivation which takes into 
account the intensity of light incident on the filter per unit angle. Here, we
provide a simple derivation to illustrate the main effect.

If $\lambda_I$ is the wavelength of a light ray 
incident at an angle $\theta$ from the optical axis (Fig.~\ref{fig:angles}), then 
from Snell's Law and Abbe's Principle,
\begin{equation}
   \left(\frac{\lambda_I}{\lambda_N}\right)^2 = 1-{{\sin^2 \theta_I}\over{n_o^2}}
\label{eq:tilt}
\end{equation}
for which $\lambda_N$ is the wavelength transmitted at normal incidence by a 
substrate of mean refractive index $n_o$.  We consider how the centroid of the 
filter passband varies in a converging beam as a function of \FR. 

For the maximum ray angle,
\begin{equation}
  \mu_o = (2{\cal F})^{-1} \sqrt{4{\cal F}^2 - 1 } .
\end{equation}
where $\mu = \cos\theta$ ($0.5 \leq {\cal F} < \infty$).
If we define $\delta\lambda = \lambda_I-\lambda_N$, the
centroid of the circular filter will vary according to
\begin{equation}
\label{integral0}
  \langle\delta\rangle = {1\over\Omega} \int_\Omega
  {{\delta\lambda}\over{\lambda_N}} \ d\Omega .
\end{equation}
Strictly speaking, equation~\ref{integral0} is only valid for small off-axis
angles.  
The fractional change in wavelength for a single ray is
\begin{equation}
  \label{ray_shift}
  \delta(\mu) =  {1\over n_o}\sqrt{\mu^2-1+n_o^2} - 1 ,
\end{equation}
in which case
\begin{equation}
\label{integral1}
  \langle\delta\rangle = {{1}\over{\mu_o-1}} \int_1^{\mu_o} \delta(\mu)\ d\mu .
\end{equation}
The quantity $\delta(\mu)$ is always negative demonstrating that the
passband shifts to bluer wavelengths.  Equation~\ref{integral1}
provides the mean shift in the filter response for an {\it on-axis}
ray bundle in a converging beam ($\mu_o \neq 1$). But the actual {\it
  range} in centroid shifts is given by equation~\ref{ray_shift} when
$\mu = \mu_o$. For example, the H$\alpha$  filter for the
UKST f/2.48 beam ($n_o = 1.97$) has a maximum centroid shift from
centre to edge of 0.54\%$\lambda$ which is a significant effect for a ${\cal R} = 100$
filter.  We now determine the mean centroid shift {\it on axis} and
verify through simulation.

The formal solution to equation~\ref{integral1} follows:
\begin{equation}
\langle\delta\rangle = {{1}\over{2(\mu_o-1)}}\left(1  - \mu_o  +
      \mu_o\,\delta(\mu_o) + {{n_o^2-1}\over{n_o}}
  \ln \left[ {\mu_o + n_o\, (\delta(\mu_o)+1)}\over{1+n_o} \right]\right)
\label{mean_delta}
\end{equation}
Over the range (slow$\rightarrow$fast)
${\cal F} \in (\infty,1)$, $\langle\delta\rangle$ varies from 0 to about 2\%$\lambda$ ($n_o \approx 1.9$).

In Appendix A, we show the general form for deriving higher order moments.
In particular, the variance of the distribution is
\begin{eqnarray}
  \label{variance}
  \langle\sigma^2_\delta\rangle =
     {{(\mu_o-1)(\mu_o+2)}\over{3 n_o^2}} - 2\langle\delta\rangle
  - \langle\delta\rangle^2 \: .
\end{eqnarray}
but this does not include the intrinsic width of the filter under consideration. 
If the filter has a top-hat profile with passband $\Delta\lambda_F$, the effective 
profile in the converging beam can be written
\begin{equation}
  \label{quadrature}
  (\Delta\lambda^\prime_F)^2 =
  \langle\sigma^2_\delta\rangle\,\lambda_N^2 
+ \Delta\lambda_{\scriptscriptstyle F}^2/12.
\end{equation}
But this expression disguises when the top hat profile is transformed
to a triangular response with the same peak transmission and passband
measured at FWHM (see Fig.~\ref{fig:on-axis}), but lower overall transmission. 
To recognize such cases requires a measure of kurtosis (see Appendix A).

In Fig.~\ref{fig:on-axis}, we simulate how a top-hat passband varies for four beams: f/1
(superfast), f/2 (Subaru Prime), f/2.48 (UKST) and f/3.3 (AAT Prime). The initial passband
is a top-hat filter with $\lambda_N =6590$\AA, $\Delta \lambda_F =65.9$\AA\ (1\%$\lambda$)
and $n =1.9$.  Both the centroid shift and passband broadening are in accordance with
the values predicted from equations~\ref{mean_delta} and \ref{variance} (see 
Table~\ref{tab:degrade}).  In beams with focal ratio f/2.5 or slower, the
on-axis passband broadening is less than 15\%\B, where \B\ is the
spectral passband in a collimated beam.
In an f/1 beam, the 1\%$\lambda$ filter broadens by a
factor of more than three with a resulting loss in average transmission. 
The band degradation is more severe for off-axis rays.

We now consider how the transmitted passband varies over the imaged
field of view. Consider the ray bundle illustrated in
Fig.~\ref{fig:rays}. With respect to the optical axis, the off-axis
rays have a skewed distribution although not as exaggerated as shown
here. If the integration in equation~\ref{integral0} is performed
from an off-axis field angle $\theta_f$, measured at the detector, to
$\theta_o-\theta_f$ ($\theta_o > \theta_f$), then
\begin{equation}
  \label{solid_angle}
  {\rm d}\Omega = -2\pi\varepsilon\; {\rm d}\mu
\end{equation}
where $\varepsilon = (\theta_o+\theta_f)/(\theta_o-\theta_f)$ for
small $\theta_f$. Roughly speaking, we need simply increase the
RHS of equations~\ref{mean_delta} and \ref{variance} by a factor of
$\varepsilon$ to examine the passband variation at an off-axis
position. In Fig.~\ref{fig:rays}, the ray bundle contours have an
eccentricity of $\cos\theta_o\; [\sec(\theta_o-\theta_f)+
\sec(\theta_o+\theta_f)]$/2  and an epicentricity of 
$\cos(\theta_o-\theta_f)/\cos(\theta_o+\theta_f)$, or about 0.2\% and 2\%
respectively for the UKST H$\alpha$ filter.
The influence of the off-axis ray bundle is evident from
Fig.~\ref{fig:on-axis}b. The passband shifts with a degree of skewness even further to the
blue.

For a 1\%$\lambda$ passband filter at AAT Prime Focus, the centroid shift
and the extreme ray shift are about 0.15\%$\lambda$ and 0.3\%$\lambda$
respectively, and the passband experiences broadening of about 5\%\B.
The beam broadening effect adds in quadrature with the manufactured passband,
such that bands narrower than about 0.5\%$\lambda$ need wide-field
compensation (see $\S$4) in an f/3.3 beam. This statement only holds for
high-$n_o$ interference filters; a more restrictive condition occurs in 
air-spaced and glass-cavity tunable filters because $n_o$ is lower.

\section{Tunable Etalon Filters in a Converging Beam}

For tunable etalon filters with air-spaced cavities (\eg TTF), we find an 
alternative treatment more instructive.  Once again, the resolving power
of the etalon cavity is degraded by fast beams.  For a dispersed spectrograph, 
\begin{equation}
{\cal R} = m.{\cal N}
\end{equation} 
where $m$ is the order of interference, and \Fin\ is the finesse (\ie the number of 
recombining beams to isolate a spectral element). The quantity \Fin\ measures the separation 
of two successive orders of interference, divided by the instrumental profile FWHM.
The spectroscopic resolution of the Fabry-Perot cavity is largely determined by the 
coating reflectivity of the two plates, but is often degraded by non-uniformities
in plate$+$coating flatness, and the beam aperture. The measured
`effective' finesse \FinE\ is given by
\begin{equation}
\frac{1}{{{\cal N}_E}^2} = \frac{1}{{{\cal N}_R}^2} + \frac{1}{{{\cal N}_A}^2} + \frac{1}{{{\cal N}_D}^2}
\label{eq:effin}
\end{equation}
where \FinR\ is the reflective finesse determined by the coatings, \FinA\ is the beam-degraded
aperture finesse, and \FinD\ is the defect finesse due to irregularities in the plate mirrors. 
Bland-Hawthorn (1995, Fig.~1) finds that, in order to minimize the impact of coating/plate 
defects, \FinE\ $\approx$ \FinR $\approx$ 40.

After Jacquinot (1954; 1960), by considering the solid angle $\Omega$ subtended by the 
innermost interference ring, we arrive at the important relation
\begin{equation}
\theta^2 = \frac{2}{\cal R} = \frac{\Omega}{\pi} \: .
\label{eq:R-Omega}
\end{equation}
Here, the peak intensity occurs at the origin and $\theta$ refers to the FWHM of the ring. 
It follows that the aperture (or beam) finesse is
\begin{equation}
{\cal N}_A = \frac{2\pi}{m\Omega} = \frac{8 {\cal F}^2}{m}.
\end{equation}
If \Res\ is dominated by either the beam or the reflective coating, 
\begin{equation}
{\cal R} = {\tt min}\left(8{\cal F}^2,m{\cal N}_R\right).
\end{equation}
for which ${\tt min}$ returns the minimum of the two values. 

The TTF $-$ shared between the AAT and WHT $-$ has \FinE $\approx$ 40 and $m \in (4,40)$, or 
equivalently, ${\cal R} \in (160,1600)$. The entire range is unobtainable in a beam faster 
than \FR $\approx$ 15 ($n_o=1$). Since the Cassegrain foci of both telescopes is close to f/8, to obtain
the full scanning range of the TTF, it is used with the TAURUS-2 collimated beam. Due to 
reflectance phase effects, the practical lower bound on \Res\ is higher than derived 
here (Jones \& Bland-Hawthorn 1998). These arise at low order spacings when the optical gap is
less than or roughly equal to the coating thickness, at which point the effective optical gap
is a complex property which depends on the details of the multilayer coatings. The net
effect is that the lowest orders cannot be achieved in practice.
 
\section{The tunable Lyot filter}

\subsection{Basic theory}

We now show that tunable Lyot filters are optimal for achieving monochromatic wide fields
at low resolving power. Wide-field compensation (or field expansion) can be used to 
generate a strictly monochromatic field.
The instrumental profile has much better wing suppression than the
Airy/Lorentzian profile.  Low orders of interference can be achieved through
the use of subtractive, rather than additive, paired birefringent elements. 

There exist excellent discussions on the basic principles of the Lyot filter
(Lyot 1933, 1944; Evans 1949; Beckers \& Dunn 1965; Title \& Rosenberg 1979, 1981).
These texts assume a proper understanding of how a wavefront propagates through a
birefringent medium. There are many approaches to a complete mathematical treatment,
including Jones and Stokes vectors, Jones and Mueller matrices, the Poincar\'{e}
sphere and impulse response analysis (Longhurst 1990; Hecht 1990).

Here, we attempt an intuitive approach; a rigorous treatment is given elsewhere 
(Bland-Hawthorn \& Title 2001).  In order to arrive at the low-resolution filter proposed 
here, there are four key elements: (a) interference through birefringence, (b) wide-field 
compensation, (c) wavelength tuning, (d) retardance through differenced elements.

The underlying principle of the Lyot filter is that light originating in a single 
polarization state can be made to interfere with itself. The first element is a polarizer
and forces unpolarized light into two orthogonal polarization components.\footnote{The
first polaroid results in a 50\% light loss, but it should be remembered that the
overall throughput of a prime focus Lyot filter is 30\% or better, comparable to
pre-existing tunable systems which involve optical paths with many more air-glass
interfaces.} If the polaroid is
oriented at 45\deg\ to the fast and slow axes of a birefringent crystal, one component
can be delayed with respect to the other. An exit polaroid, aligned with the entrance
polaroid, recombines the orthogonal components to effect interference.

Crystalline materials are usually, but not exclusively, birefringent in that the 
refractive index is different along two axes within the crystal.
In a positive uniaxial crystal (\eg quartz), the refractive index experienced by the 
ordinary ray on the `fast' axis, $n_o$, is lower than for the extraordinary ray, $n_e$, 
along the `slow' axis (see Fig.~\ref{fig:biref}). The birefringence $b=n_e-n_o$ 
leads to a time delay $\delta t = b d /c$ where $c$ is the speed of light in vacuo, and 
$d$ is the thickness of the element.  Retardance is expressed in angular units, \ie\ 
$r=2\pi b d /\lambda$.

For light traversing the simple system in Fig.~\ref{fig:montage}(i), the transfer function 
is given by Malthus' law (\eg Hecht 1990),
\begin{equation}
T(\lambda) = cos^2(\pi b d / \lambda) .
\label{eq:xfr}
\end{equation}
Lyot (1933) realized\footnote{The principle was independently discovered by 
\"{O}hman (1938).} that a cascading series of aligned birefringent units could be used 
to isolate a spectral element. From equation~\ref{eq:xfr}, the easiest way to arrange 
this is to make each element twice the thickness of the preceding element (see 
Fig.~\ref{fig:lyotprof}). The interleaved polaroids are essential to the rejection
of out-of-band emission. The instrumental response is magnified in 
Fig.~\ref{fig:compprof}. It is clear that  the 5-stage Lyot has much better wing
suppression compared to the Lorentzian response of the Fabry-Perot. The core of the profile
is highly triangular and closely approximates the Gaussian response.

Clearly, the free spectral range is determined by the thinnest element, and the 
spectral resolution by the thickest element. Lyot (1944) demonstrates that the
instrumental response of a series of $q$ elements is
\begin{equation}
T(\lambda) = \frac{1}{4^q} \frac{sin^2(2^q \pi b d_o / \lambda)}{sin^2(\pi b d_o / \lambda)}
\label{eq:xfrm}
\end{equation}
It follows that the free spectral range is $\Delta \lambda = \lambda^2 / b d_o$,
the finesse is ${\cal N} = 1.13\; 2^q$, and the spectral element $\delta\lambda$ is
the ratio of these two quantities. These quantities can be related directly to
equivalent parameters for Michelson and Fabry-Perot interferometers 
(Bland-Hawthorn \& Cecil 1997). 

Low orders of interference are difficult to achieve at optical wavelengths. We have
already mentioned peculiarities involving coated etalon cavities (\S 3). Now consider 
the problem involved in replacing the air-spaced cavity with a solid. This would require 
the manufacture of a micron-thick sliver of glass which was then polished to optical
quality. This is easier to accomplish with birefringent materials simply because 
much thicker materials are needed to achieve a given time delay
when compared to solid etalons.

The resolving power of air-spaced Fabry-Perot plates a distance $d_o$ apart is 
${\cal R} = 2 d_o {\cal N}/\lambda$, where \Fin\ is the reflective finesse of
the coated plates. This compares with ${\cal R} = b d_o {\cal N}/\lambda$,
where \Fin\ is now the Lyot finesse. The Lyot element with thickness $d_o$
determines the order separation (fringe spacing).

For two commonly used materials, MgF$_2$ and crystal quartz, $b\approx 0.01$, 
which means that, for the same resolving power, the thinnest element of the Lyot 
is two orders of magnitude thicker than the equivalent plate spacing of the TTF.
If our aim is to block neighbouring orders with conventional ${\cal R} \approx 5$
filters, the required TTF sub-micron spacings cannot be achieved in practice, and
are subject to reflectance phase effects (\S 3). For the
Lyot, it is possible to optically bond sub-mm thick elements to a supporting 
substrate, or to use thicker retarding elements in a subtractive arrangement ($\S$4.5 
and Fig.~\ref{fig:montage}(vi)) to 
achieve the same equivalent optical thickness.

\subsection{Wide-field compensation}

In $\S$2 and $\S$3, it was shown that a narrow spectral band cannot be isolated in a fast 
beam using interference or etalon filters. This is because the path length through the 
resonating cavity cannot be equalized for all off-axis angles. Remarkably, off-axis path 
lengths {\it can} be equalized with crossed birefringent elements (Lyot 1944; Evans 1949) 
even for resolving powers as high as \Res~$\sim$~$10^5$ (Title \& Rosenberg 1981). Wide-field 
compensation has been demonstrated to be extremely effective for narrowband imaging
by several groups, particularly in the fields of remote sensing and solar astronomy (Beckers,
Dickson \& Joyce 1975; Bonacinni\etal\ 1989).

An intuitive explanation for why the method works is hard to come by. To begin,
consider how wide-angle, zero-wave retarders are made. The retarder element in
Fig.~\ref{fig:montage}(i) is split perpendicular to the propagation axis into two
equal thickness elements. The pieces are then bonded together after rotating the
fast axis of the second element to line up with the slow axis of the first element.
There is no net retardance for any angle through the retarder. 

The principle of balancing fast and slow propagation is the primary method of 
wide-field compensation. In order to avoid the simple cancellation of the zero-wave retarder,
a half-wave plate is placed between the split retarder elements (see Fig.~\ref{fig:montage}(ii))
oriented at 45\deg\ to the fast and slow axes.  Why this works is somewhat involved: the most
accessible discussion is given by Title \& Rosenberg (1979).

The half-wave plate advances the phase of the $o$-ray by 180\deg\ (see Fig.\ref{fig:halfwave}) 
with respect to the $e$-ray. This rotates the plane of polarization by 90\deg\ and re-aligns the
$o$-ray with the fast axis of the second element. Hence, the {\it on-axis} time delay of 
the crossed elements (Fig.~\ref{fig:montage}(ii)) remains {\it unchanged} from the original system 
(Fig.~\ref{fig:montage}(i)). It is the {\it off-axis} rays which benefit in two respects.

While the off-axis behaviour of interference filters
is isotropic (circular isochromes), uniaxial crystals have a strong azimuthal dependence 
(hyperbolic isochromes). For rays at an arbitrary azimuthal angle $\phi$
(see Fig.~\ref{fig:angles}), the general form of the retardance is 
\begin{equation}
r(\theta,\alpha) = r_o \left[ 1 - {\frac{\sin^2 \theta}{2 n_o}} \left({\frac{\cos^2 \phi}{n_o}} - {\frac{\sin^2\phi}{n_e}} \right) \right]
\label{eq:retard}
\end{equation}
Note that the retardance changes sign between neighbouring quadrants, and this is the physical basis
for wide-field compensation. Remarkably, with the half-wave plate in place, the complex azimuthal
behaviour all but disappears.  Light which enters the first split element from the direction 
$(\theta,\phi)$ enters the second element from the direction $(\theta,\phi+90$\deg). From 
eq.~\ref{eq:retard}, the overall retardance is now
\begin{equation}
r = r_o \left[ 1 - {\frac{\sin^2\theta}{4 n_o}} \left({\frac{n_e-n_o}{n_o}}\right) \right]
\label{eq:wide-field}
\end{equation}
The equivalent form for the interference filter is (eq.~\ref{eq:tilt})
\begin{equation}
r = r_o \left[ 1 - {\frac{\sin^2\theta}{2 n_o^2}} \right]
\label{eq:filters}
\end{equation}
The monochromatic, solid acceptance angle available to a compensated birefringent filter 
is a factor of $2 n_o/(n_e-n_o)$, \ie orders of magnitude, larger than possible with an 
interference filter.
Moreover, Lyot filters have a much greater ${\cal R}.\Omega$ product than available to
interference filters and etalon filters (eq.~\ref{eq:R-Omega}).

\subsection{Tuning}

Title \& Rosenberg (1981) demonstrate several schemes for tuning a Lyot filter.
By far the best scheme involves a quarter-wave plate.  The reader should consult
Evans (1949) for the most accessible explanation of how tuning works.

The output of a given birefringent element is composed of equal
amplitude rays along the fast and slow axes, with a relative phase
shift which depends on the wavelength. The output can also be
described as an elliptically polarized beam, \ie\ two rays of unequal
amplitude with a 90\deg\ phase shift between them.  The ellipse is aligned with the
entrance polarizer and the ellipticity depends on the wavelength. The
trailing quarter wave plate removes the 90\deg\ phase shift and
transforms the elliptically polarized light to linearly polarized
light.	The orientation $\psi$ (and therefore wavelength) is a function 
of the ellipticity $\varepsilon$ such that
\begin{equation}
\psi = \tan^{-1} \varepsilon
\end{equation}
where
\begin{equation}
\varepsilon = -\tan \frac{\pi d b}{\lambda} .
\end{equation}
The tuning relationship for a single stage reduces to 
\begin{equation}
\psi = \pi d b (\frac{1}{\lambda} - \frac{1}{\lambda_o})
\label{eq:psi}
\end{equation}
where the natural wavelength $\lambda_o$ is selected at $\psi = 0$. In practice,
tuning is achieved by rotating the exit polarizer.
The exit polarizer of one stage constitutes the entrance polarizer of the next
stage. Peak transmission is achieved as long as each successive
stage is rotated precisely twice the rotation of the preceding thinner element.
A gear system which maintains 2:1 synchrony between successive stages
presents an interesting mechanical problem, particularly when anti-backlash
gearing is incorporated (\eg Beckers \& Dunn 1965).
The ($\psi,\lambda$) relation for the complete Lyot system over the full optical 
range is more involved than equation~\ref{eq:psi} (Beckers, Dickson \& Joyce 1975).
Conveniently, the sensitivity of the wavelength tuning to angular errors decreases 
for the thicker elements. 

For the AAT design, we adopt a more convenient form of opto-mechanical control
than conventional gearing (Beckers \& Dunn 1965). Each of the Lyot stages is
rotated independently and accurately with the use of circular encoders and 
separate motors.  This has several advantages. No element needs to be rotated more 
than $\pi$ radians which reduces the heat input due to the fastest rotating elements 
in the conventional design.  The GSFC Lyot filter requires 400 rotations of the thickest
element to tune over the full optical range (Palunas\etal\ 2000).

Furthermore, alignment errors between the Lyot stages can 
be calibrated and compensated for.  
As each element rotates, it causes a cosine-squared modulation of the intensity at a 
single wavelength (eq.~\ref{eq:xfr}). Since each element produces its own distinctive
modulation, it is straightforward to align the transmission maxima with a laser.
Moreover, each element can be calibrated separately for thermal variance
in a temperature-controlled environment. 
If we connect each element to a separate temperature probe, simple 
temperature adjustment can be applied independently from a software table. 

\subsection{Manufacture \& construction}

\noindent {\bf Materials.}
A wide-band Lyot requires a uniform response in wavelength. In order of optimal
birefringence, the best substances are MgF$_2$, sapphire and crystal quartz
(Serkowski 1974).  All of these produce roughly comparable retardances, \ie 
$\vert b \vert \approx 0.01$.  
Calcite has an order of magnitude higher birefringence with natural cleavage planes.
However, it is brittle and difficult to handle in thin slices.
Sapphire is very hard and therefore requires diamond tooling and very long polishing
times. At the other extreme, LiNbO$_3$ and KDP are hygroscopic and much too pliable.
Crystal quartz and MgF$_2$ are hard, readily available and reasonable
to work with. While quartz is the easier to work with,  MgF$_2$ has the most
achromatic birefringence of optical crystals. But the lack of natural cleavages means
that both materials must be cut and polished.
Since natural crystals are mined, it is hard to come by large pieces with high
levels of purity. Instead, most crystal elements are made from synthetically grown
materials.  
The required crystal dimensions of 160~mm for MgF$_2$, sapphire and crystal quartz
are mildly challenging for elements of sufficient purity, but not without precedent 
(Title 1999).

\noindent {\bf Thin elements.}
A practical concern is producing the thinnest elements which drive the free spectral
range. A primary aim is to use conventional filters ($m = \lambda/\Delta\lambda 
\approx 5$) for blocking neighbouring orders, the thickness of the thinnest element
is $d_o = m\lambda/b$, or about 0.2~mm for crystal quartz, sapphire, and so on.
Hariharan, Oreb \& Leistner (1984) have demonstrated that large elements can be
`float' polished down to thicknesses of 0.15~mm quite readily. The retarding material
is bonded to a neutral substrate before polishing.  The polishing can be controlled to 
a thickness tolerance of 1\% or better which is the required accuracy for our Lyot
filter design. After polishing, the element can be strengthened by leaving it bonded 
to the substrate.

There are other approaches to the thin element problem. Firstly, we could choose to 
make the thinner elements with a lower birefringence material (composite Lyot filter). 
Second, to construct the zero-wave retarder in $\S$4.2, a single element with 
$r= (n_e - n_o) d_o$ was adapted to $r = A(n_e - n_o)d_o + B(n_o - n_e)d_o$ where $A=B=0.5$.
Therefore, we can advance, retreat or cancel retardance along a given axis by a 
suitable choice of $A$ and $B$. Setting $A=3$ and $B=2$, for example, provides 
equivalent retardance to a single element with $r = (n_e - n_o)d$, illustrated in Fig. 6(vi).
Thus, thicker
elements used together can be used to replace a single thin element. However, these
may require wide-field compensation.

\noindent {\bf Polaroid transmission.}
Polaroid filters are notoriously lossy with typical transmissions of 70\%.
However, thin film polarizers made by `bleaching' can have transmissions 
as high as 98\% (Gunning \& Foschaar 1983). 
A polymerizing sheet is placed in acetone to dissolve most of the plastic. 
The polarizing film is then stretched and deposited over a substrate in
a humid oven. Much of the art of bleaching resides in technical reports of the
Lockheed Palo Alto Research Laboratory where it was first developed by H.E.~Ramsey.
When achieving the highest transmissions through bleaching, it is important that the 
extinction ratio (the ratio of the blocked and transmitted light) be kept to 
10$^{-3}$ or better. 

\noindent {\bf  Wave plates.}
The major technical challenge of the wide-field Lyot filter is to produce achromatic 
half-wave and quarter-wave plates of sufficient size and quality (Serkowski 1974).
There are different levels of sophistication
involving either one, two or three sandwiched plates of birefringent material.
The degree of difficulty and cost increases dramatically with the number of
elements. A single birefringent retarder can really only operate at a single 
wavelength since $b(\lambda)$ varies by 10\% or more over the optical band (Serkowski 1974).
Although a single retarder element with $R\sim 150$ would be useable over at least 50~nm
at \Hb.  Large monolithic elements can be made and are relatively inexpensive.  In terms 
of wavelength invariance, MgF$_2$ has almost ideal properties, but is harder to work with 
than crystal quartz.

A wave-plate with better achromatic behaviour combines two plates of different birefringent 
materials, in particular, quartz and MgF$_2$ (Clarke 1967). These plates were used in the 
Zeiss universal filter (Beckers\etal\ 1975) optimized for $R=30,000$ (at \Ha) over the
range 450-700~nm. Two-layer composite waveplates larger than 100~mm have been 
made but are costly and constitute a significant technical challenge. 
The best achromats exhibit variations in retardance of tens of degrees. 

The universal filter developed by Lockheed Solar Observatory achieved $R=50,000$ 
(at \Ha) over a slightly broader response using waveplates comprising oriented sheets of 
polyvinyl alcohol (Title 1999). These are polymer films stretched to align the long-chain,
organic molecules much like cellophane. The polarization is produced by electrons 
within the lattice which experience different binding forces between the parallel and 
perpendicular directions.  The advantage of cellophane is that arbitrarily
large elements can be made although the transmission is not as good as conventional
retarders.

A three-element 
retarder\footnote{Fractional wave retarders can be generalised at great expense to at 
least 10 birefringent elements (Harris \& McIntyre 1968; Title 1974).}, 
\ie the super-achromatic retarder (Pancharatnam 1955), has excellent broadband properties, 
with variance in retardance of only a few degrees, but rarely exceed 50mm in diameter and 
are very expensive to make.  

Bigger retarders can be made by mosaicing smaller elements. For an AAT Lyot filter
with a 30\arcmin-40\arcmin\ field using commercial wave plates, this would require four 
elements to cover the field in the case of the achromat, and nine 
elements for the superachromat. This is a highly undesirable arrangement for an imaging
system. The supporting structure typically has $1-2$mm wide frames
and it becomes difficult to accurately align the individual elements. This
approach is more relevant to Lyots placed at the pupil in a collimated beam
(see Fig.~\ref{fig:nonuniform}).

An entirely new prospect is the possibility of sub-lambda gratings in transmission
(Kikuta, Ohira \& Iwata 1997).
These have the potential to yield close to 100\% transmission over a broad spectral
region. Artificial optical anisotropy can be produced in a {\it homogenous} optical
material when the surface of the material is covered with a regular grating whose
period is smaller than the wavelength of light. The transmitted light will be delayed
according to the orientation of the $\vec{E}$ vector to the ruling. With recourse
to effective medium theory, the refractive indices can be expressed in terms of the
groove depth $d$ and groove frequency $q$,
\begin{eqnarray}
n_e &=&   \sqrt{n_r^2 q + n_g^2 (1-q)} \\
n_o &=& 1/\sqrt{\left({1\over n_r}\right)^2 q + \left({1\over n_g}\right)^2 (1-q)} .
\end{eqnarray}
The indices $n_r$ and $n_g$ are for the grating ridges and grooves respectively.
At a wavelength of 1$\mu$m, a line width of 250~nm and groove depth of 800~nm
could be made to operate like a conventional achromat. These lines are cut with
either holography or e-beam lithography combined with reactive ion etching.  The 
advantage here is that much larger elements are possible and the grating can be 
much thinner in the direction of the optical axis compared to commercial achromats.

\noindent {\bf Angular alignment.}
Alignment of the separate Lyot stages is simplified by the the optomechanical control 
described in \S4.3. In practice, the crystal optic and polarizer axes can be 
matched to better than 30\arcmin\ which is adequate for the Lyot described here.
Within a single Lyot stage, the tolerances for registering the crystal optic
and polarizer axes are rather stringent (Steel, Smartt \& Giovanelli 1961) but are 
not a major challenge. Accuracies of 6\arcmin\ or better can be readily achieved 
(Giovanelli \& Jefferies 1954).  Commercially available wave plates constructed 
from sandwiched retarder elements can be internally aligned to better than 1\arcmin.

\subsection{Expected system performance \& design requirements}

\noindent{\bf Throughput.}
The Goddard Space Flight Center (GSFC) Lyot filter incorporates eight tunable
stages, and has a clear diameter of 100mm and an optical path length (thickness) of
$n_o \ell = 375$~mm (Palunas\etal\ 2000). Our design below is based on the five lowest
order elements of this system, first proposed by Beckers \& Dunn (1965).
B.A.~Gillespie has measured
the GSFC Lyot filter as having better than 30\% throughput to unpolarised light.
For an optical path length of $n_o \ell = 145$~mm, we can expect a somewhat higher
transmission of 40\% or better for the low resolution filter. The modified doublet
is expected to have a throughput close to 90\% over the optical range. The WFI
detector comprises 8 MIT-LL 4K$\times$8K CCDs. We calculate an overall system
throughput of 32\% to unpolarized light ($\lambda$750~nm) with the WFI mosaic CCD,
comparable to the existing TTF system (Jones 1999).

\noindent{\bf Image quality.}
To ensure the quality of the optical wavefront, the retarder elements are polished
to $\lambda/4$ per inch.  The flatness criterion is an rms average over the element 
area and is straightforward to achieve with modern polishing techniques. This
assumes there are no high frequency ripples and that the material optical homogeneity 
is at least several parts per 10$^5$ or better.

The Universal Birefringent Filter (Beckers\etal\ 1975) has 58 precision crystal 
elements and polarizers
in series, within an optical system comprising four compound lenses, four prisms,
a Fresnel rhomb, a thin-film filter, two external polarizers, two mica waveplates,
and two KD$^*$P modulators. (This is three times the number of elements for the
low order Lyot system discussed here.) Smartt (1979) used a point diffraction 
interferometer to demonstrate that both the on- and off-axis performance give
good wavefront uniformity.

\noindent{\bf Spectroscopic integrity.}
In order to achieve the theoretical bandpass, the faces of the retarder element must
be parallel. Each crystal element of the Lyot acts as a two-beam interferometer.
However, the manufacturing tolerances are much less stringent than for the Michelson 
interferometer. If the shift in central wavelength is to be no more than a fraction $p$
($0<p<1$) of a resolution element, the crystal faces must be parallel to
\begin{equation}
{\cal P} = {p\over{(n_e - n_o)}} \left({\lambda \over 2}\right) .
\end{equation}
In other words, the criterion is set by the time delay rather than the physical or
optical path length.
For crystal quartz, magnesium fluoride and sapphire, assuming $p=0.01$, this amounts to 
no more than $\lambda/2$. In practice, this is somewhat stringent although this is 
probably a worthwhile target to counteract the cumulative effect of 5 elements in series.
Indeed, the requirement of good image quality (see above) already assures the
spectroscopic integrity of the Lyot profile.

\noindent{\bf Scattered light.}
The quoted transmissions assume sol-gel $+$ MgF$_2$ coatings on all external
surfaces. The mean anti-reflectivity is better than 99\% over the optical window
380~nm to 1600~nm.  While the sol-gel layer is porous, after a simple DDMS treatment,
it repels moisture very effectively (Stilburn 1999; Thomas 1999). Since the coatings 
are external to the polarizers, the polarizability of the coating layers has negligible impact
on the Lyot operation (see $\S$ 5). 

To avoid internal scatter, 
all elements of the Lyot are immersed in index matched oil. If the 
molecular chains are too long, the higher mean molecular weight tends to
produce higher viscosity. If the chains are too small, the oil
has the tendency to produce heat through turbulence.  Dow Corning
and Mobil manufacture synthetic lubricants (\eg silicone oil)
with chain lengths designed to encourage a smooth flow,  \eg Dow Corning 
silicone oil 702 with a refractive index, $n_i = 1.512$. The oil must have low 
viscosity to avoid thermal input from the mechanical tuning, or the temperature
of the Lyot chamber must be actively controlled (\eg Steel, Smartt \& Giovanelli 1961).
For quartz, the shift is $-$0.5\AA\ K$^{-1}$ in the visible. A full discussion of
thermal influence on the phase constancy is given by Lyot (1944).
For the low resolution filter envisaged here, temperature control is not essential
(although monitoring may be desirable). 

\noindent{\bf Wavelength calibration.}
The wavelength dispersion over the full optical range is expected to
be fairly linear. Palunas\etal\ (2000) shows that the GSFC Lyot requires 400 
rotations to cover the 400$-$700 nm range. The spectral window is accurately 
calibrated using a quintic curve (with a dominant linear term) to fit 28 Ne lines 
to an rms scatter of 1.2\AA.

\section{The Low Resolution Lyot Filter} 

\subsection{Overview}

We now describe the particular structure of our proposed system for the AAT f/3.3 Prime Focus.
Our design comprises quartz elements and makes use of stages (iii), (iv) and (vi) of 
Fig.~\ref{fig:montage}. Only the two or three thickest elements require wide-field 
compensation. A 5-stage Lyot filter produces a natural finesse of ${\cal N}=36.2$.
For the order sorting filters,
we choose to adopt the revised Gunn filter bands (Gunn\etal\ 1998):
356/60, 483/138, 626/138, 767/154, 910/137 (central wavelength/bandpass in nm). 
If we set the free spectral range $\Delta\lambda$ to 138~nm at 550~nm,
the basic unit is a retarder thickness $d_o\approx 0.22$~mm (Fig.~\ref{fig:lowres}).
The periodic response of the Lyot filter (${\cal R} = 144$) is then properly blocked at all 
wavelengths by the Gunn bands.

This a particularly convenient choice for several reasons. The revised Gunn bands have
high transmission, and are very square, which makes them well suited to blocking neighbouring 
orders.  Within a few years, the Sloan Digital Sky Survey (Gunn\etal\ 1998) will become the 
accepted photometric system providing high quality data and photometric standards with 
small differential errors over most of the sky.
The Gunn filters incorporate multilayer dielectric coatings in order to achieve the 
fast cut-off to low energies. But the broadband filters are external to the first 
polaroid and therefore do not affect the performance of the Lyot filter.
Our optical design in Fig.~\ref{fig:optics} (see Table~\ref{tab:spots}) ensures excellent 
imaging performance over the full field through both the Lyot and Gunn filters.

\subsection{Optical design}

The AAT has Ritchey-Chr\'{e}tien optics with a hyperboloidal primary mirror.  This requires 
correction at prime focus for spherical aberration and coma.  Three prime focus correctors 
already exist: a single element silica aspheric (10 arcmin), a silica all spherical doublet 
(25 arcmin) and a UBK7 all spherical triplet (60 arcmin), where the unvignetted field is 
given in brackets (Wynne 1972, 1974).

Considering the low order Lyot system as a plane parallel plate of silica 90 mm thick, it 
introduces spherical aberration of the same sign as that from the hyperboloidal mirror but 
of much smaller amount.  If the plate were to be preceded by a perfect imaging system with 
focal ratio f/3.2 (the focal ratio applying with the doublet corrector), its spherical 
aberration would result in an rms image diameter of 0.34 arcsec.  

Using the Zemax ray-tracing program, we have examined the performance of the existing triplet 
and doublet in conjunction with a 6 mm thick filter and with the low order Lyot system.  
The optimum position of the detector surface in relation to the corrector lens changes considerably 
between the two cases.  In assessing performance, this back focus has been fixed for each of the 
two cases but the position of the corrector lens in relation to the primary mirror has been 
reset to an optimum for each wavelength or wavelength band considered.  This freedom is readily 
available on the AAT, where the fine focus adjustment is provided by translating the whole telescope 
top end.

Figure~\ref{fig:optics} (top) shows the layouts of the two systems for the two configurations. For 
the doublet, 
performance was assessed over its unvignetted field of 25 arcmin and for the triplet, over 40 arcmin 
diameter - the most that we considered to be practical with the Lyot system. For imaging through 
the Lyot filter, five single wavelengths at the centres of the Gunn passbands were considered.  
For broader band imaging, each of the Gunn passbands was represented by three wavelengths.  
Table~\ref{tab:spots} lists the image sizes found for the various spectral cases and for three fields: 
on axis, the extreme field radius, and 70\% extreme radius.  Cases where the rms image diameter
is 0.4 arcsec or more are underlined. The doublet performs very well other than being limited 
to 25 arcmin field diameter.  Evidently, shifting the doublet axially from its standard relationship 
to the primary mirror has allowed significant compensation of the spherical aberration introduced 
by the 90 mm thick filter.  The narrow band images with the triplet are worse than is desired, 
especially at the long wavelength end of the range. This is not surprising since the triplet's 
designed range was $B$ through to $V$.

The high image quality with the doublet encouraged exploration of a doublet design with enlarged 
field. Figure~\ref{fig:optics} (bottom) shows the layout of the design which resulted, having 40 
arcmin diameter 
unvignetted field and using UBK7 glass for both elements.  Higher weighting was given in the 
design to the very narrow band imaging through the Lyot filter than to the broader band imaging. 
Other than for imaging through the two shorter 
wavelength Gunn filters, its performance is very good.  An alternative design with silica elements 
rather than UBK7 has performance a little inferior to that with UBK7.  With its total thickness in
the doublet of about 30 mm, the internal transmission of this glass is about 77\% at 330 nm, 93\% at
350 nm, and 99\% at 400 nm.

\section{Discussion}

To date, it has not been possible to undertake truly monochromatic wide-field, 
narrowband imaging since air-gap tunable filters and monolithic interference filters 
undergo severe bandpass degradation (\S 2).  A wide-field Lyot filter at Prime Focus 
has important astrophysical applications for a wide range of astrophysical problems, 
which we now discuss.

\subsection{Star formation at $z\sim 1$ and photometric redshifts}

Wide-field tunable filters offer a major advantage for
targetted redshift surveys compared to all conventional 3D 
spectrographs currently operating or planned for the 8-10m class
telescopes. Put simply, it is extremely difficult to format spectra
over a wide-field CCD mosaic. 

Tunable filters are now routinely used to obtain a series of narrowband 
images over a wide field.  Examples of \Ha\ line detections from
compact star-forming galaxies are shown in Fig.~\ref{fig:collage}. The 
postage stamps are taken from a series of 
10\arcmin\ images in the {\sl TTF Field Galaxy Survey} (Jones \& Bland-Hawthorn 
2001).  In targetted surveys, a narrow redshift 
interval tuned to, say, the redshift of a cluster is compared to a redshift 
interval offset from the cluster.  Also, star forming galaxies are
easily detected in [OII]$\lambda$3727 at $z\sim 1$ (Baker\etal\ 2001). 
The corresponding flux levels for a given star formation rate for both
\Ha\ and [OII] are shown in Fig.~\ref{fig:SFR}.  

Our expectation is that large-scale (supercluster) structure should be 
visible at $z\sim 1$ on 30\arcmin\ scales. In Fig.~\ref{fig:GIF}, we show the 
results from a $\tau$CDM simulation by the GIF consortium (Kauffmann\etal\ 1999).
The simulations adopt ($\Omega_m=0.8$, $\Omega_{\Lambda}=0.2$, $h=0.5$) and are 
normalised to the present day cluster abundance ($\sigma_8 = 0.6$).
The simulation shows the distribution of galaxies with star formation
rates in excess of 0.3 M$_\odot$ yr$^{-1}$ (LMC). This is our 4$\sigma$
sensitivity limit to [OII] in 1\arcsec\ seeing. 
The white points are what would be visible in a single 4~hr image 
for $R\sim 500$. The black points show the additional objects from
lowering the resolution to $R\sim 150$. In the same exposure time, 
the Lyot's sensitivity to LMC star-formation rates at $z=1$ is now 2.5$\sigma$.

Once emission line sources are detected, the next step is to establish
that we have identified the emission line correctly. If the instrument
is sensitive to $m$ other emission lines intrinsic to the survey
objects, the survey slice is sensitive to $m$ other redshift intervals.
Redshift ambiguity affects broadband-selected (\eg Lilly\etal\ 1995)
and emission-selected samples alike (\eg Yan\etal\ 1999).

In Fig.~\ref{fig:phot-z}, we explore the
use of broadband $UBRI$ colours as a means of solving the redshift 
ambiguity inherent in tunable filter detections of a single line.
Our use of the photometry is not intended to provide photometric redshifts 
in the classical sense (cf. Koo 1999; Fukugita, Shimasaku \& Ichikawa 1995).  
The volume-limited nature of 
the narrowband search {\sl necessarily} limits emission-line candidates
to a few specific redshift slices.  This overcomes a major hurdle for 
conventional photometric redshift methods (cf. Brunner\etal\ 1997) in that 
a number of potential redshift solutions is known at the beginning.
We only require that the broadband colours be able to distinguish
{\sl between} these different redshifts, which requires that they are 
sufficiently well-separated in colour-colour space. At the wavelength 
intervals scanned for the {\sl TTF Field Galaxy Survey} $-$ 668/21, 
707/26, 814/33, 909/40~nm $-$ the optical emission lines useful for
measuring star-formation rates (\Ha, \Hb, \OII) are separated in
redshift by $\delta z = 0.3$ or more. 
We cannot distinguish \Hb\ from [OIII] using phot-z estimates, 
nor can we distinguish \Ha\ from [NII]. But in most cases,
these are not normally confused since the spectral scan will 
reveal paired lines if [NII] is strong, or a doublet if we
are looking at [OIII] as opposed to \Hb.

For the {\sl TTF Field Galaxy Survey}, colours from the galaxy spectra used
by Fukugita\etal\ (1995) were re-computed for us by Dr.~K.~Shimasaku for the 
emission-line redshifts that fall within the TTF filters.  Figure~\ref{fig:phot-z} 
shows the distribution of galaxy colours objects obtained in four passbands 
($UBRI$). There is a single track derived for each emission line at the 
required redshift in Table~\ref{tab:lines}. Along each track, the change 
in colour due to galaxy type is 
spanned from Im to E galaxies.  The combination of $B-I$ and $U-R$ colours 
was found to present the most widely separated tracks followed
by $-$ in order of decreasing usefulness $-$ combinations of $UBI$, $BVRI$, 
$BRI$.  The colour tracks are computed from the Fukugita\etal\ spectra
within the blocking filters used in the {\sl TTF Field Galaxy Survey}.

Tunable filter surveys are at a particular advantage in observations
of fields with pre-existing deep broadband data. If we are able to
establish accurate phot-z values for most targets in the field,
this increases our volume advantage by roughly $m$ since the wavelength
interval corresponds to $m$ other redshift intervals. Over the next
few years, large areas of sky are to be mapped in deep imaging surveys
(\eg\ McMahon\etal\ 2000) down to at least $m_{\rm AB} = 26$ in the 
major broad bands. These fields are a natural choice for future wide-field
Lyot observations.

\subsection{Absorption line sources}

The TTF has been used with limited success to observe absorption-line sources.
The main limitations were the highly Lorentzian instrumental profile and the phase 
variation over the field. The Lyot profile has much better wing suppression than the 
TTF profile (see Fig.~\ref{fig:compprof}). This is particularly favourable for absorption
line studies of stellar fields or extended continuum sources. Furthermore, the proposed
design would be fully monochromatic to much better than 1\AA\ variation over the field.
This could benefit any number of wide-field, stellar absorption line studies:

\begin{enumerate}

\item Identification of clusters from k+a galaxy concentrations (Dressler \& Gunn 1983; 
Jones 1999).

\item Use of integrated stellar spectral indices in ellipticals, galactic bulges
and globular clusters (Faber 1973; Mould 1978; Worthey, Faber \& Gonzalez 1992; 
Jablonka, Martin \& Arimoto 1996).

\item Measurement of metallicity gradients in galaxies using Mg$_2$ and Fe (527~nm) 
from the evolved stellar population (Molla, Hardy \&  Beauchamp 2000). To date, 
most of the abundance information on spirals has come from $\alpha-$processed elements
using HII region emission line diagnostics.

\item Measurement of chemical inhomogeneities in dwarf galaxies from the
CN (388nm) and Ca~II~H+K bands (Smith \& Dopita 1983).

\item  Discrimination of carbon from M stars: 780nm is in a TiO band for M stars 
but is essentially continuum for C stars; 810nm is in a CN band for C stars and
basically continuum for M stars. A plot of $780-810$ vs. $V-I$ is a good discriminant
between these stars (Wing \& Stock 1973; Cook, Aaronson \& Norris 1986; 
Cook \& Aaronson 1989; Richer, Crabtree \& Pritchet 1990).

\item Discrimination of giants from dwarf stars: the latter have stronger MgH (526~nm) 
at the same temperature than giants (Clark \& McClure 1979).

\item Measurement of stellar metallicities from narrowband imaging at Ca~II~H+K and
Ca~II triplet at 870nm (\eg\ Anthony-Twarog \& Twarog 1998).

\end{enumerate}

\subsection{Extended line sources}

The literature on extended emission-line sources covers a broad range of topics. 
The following list is representative and certainly not exhaustive:

\begin{enumerate}
\item Identification of high-redshift Ly$\alpha$ galaxes (Lowenthal\etal\ 1991;
Francis\etal\ 1996; Stiavelli\etal\ 2001).

\item Detection of warm intra-cluster gas in nearby groups of galaxies
(Fukugita, Hogan \& Peebles 1998; Maloney \& Bland-Hawthorn 1999).

\item Detection of warm gas in extended radio sources which can subtend huge angular scales, 
\eg Cen A ($>$10\deg; Junkes\etal\ 1993). Radio lobes (\eg Cyg~A, For~A) 
frequently exhibit large-scale, highly structured depolarized regions (Fomalont\etal\ 1989)
which in For~A are clearly associated with warm gas (Bland-Hawthorn\etal\ 1995).
Ionized hydrogen may well be associated with relic radio lobes and
radio haloes (\eg M87; Owen, Eilek \& Kassim 1999).

\item Extended gas in cooling flow clusters (Voit \& Donahue 1999; Voit, Donahue \& Slavin 1994;
Jaffe \& Bremer 2000).

\item Extended optical emission associated with jets from active galactic nuclei
(\eg Cen A; \qv Israel 1998). This includes ionization cones and ionized plumes which have 
now been detected to the HI edges of some Seyfert galaxies (Shopbell\etal\ 2001).
Shopbell\etal\ (1999) have identified a very extended nebula around an x-ray selected QSO.

\item Galaxy-scale bowshocks from the motion of a galaxy through a cluster or a
group medium (Stevens, Acreman \& Ponman 1999; Veilleux\etal\ 1999).

\item Line emission associated with starburst winds extending over large scales, \eg M82 
(Devine \& Bally 1999).

\item Dynamics of diffuse stellar structures, the outer parts of galaxies and the 
intracluster medium using free-floating planetary nebulae (Theuns \& Warren 1997).

\item Tracing star formation in spiral (Rozas, Beckman \& Knapen 1996) and
dwarf (Stewart\etal\ 1999) galaxies.

\item Diffuse ionized gas in the Galaxy (`Reynolds layer') and nearby galaxies,
\eg M31, M33, LMC (Dettmar 1992; Dahlem 1997; Haffner, Reynolds \& Tufte 1999).  
Moreover, \Hb\ emission is notoriously difficult to image in spiral 
galaxies due to the underlying absorption from the stellar population. There now 
exist differential imaging techniques which allow for an automatic correction
for the underlying absorption (Cianci\etal\ 2000).

\item The Magellanic Stream \& Magellanic Bridge have been detected in \Ha\
at isolated positions (Marcelin, Boulesteix \& Georgelin 1985; Weiner \& 
Williams 1996). A complete map of the \Ha\ distribution would require an 
extensive campaign with a wide-field narrowband imager.

\item A vast array of emission nebulae associated with compact sources, 
including photo-dissociation regions, supersoft x-ray sources, x-ray binaries,
potassium shells around stars, planetary nebulae, and supernova remnants.
Extended nebulae are seen around fast-moving pulsars and anticipated around soft 
gamma-ray repeaters and gamma ray bursts. Tunable filters are also well adapted
to constructing emission line source catalogues (Murphy \& Bessell 2000).

\item The use of broadband colours and \Ha\ emission in studies of young star
clusters to identify brown dwarfs (Tinney 2000) and Be stars (Keller\etal\ 2001).

\end{enumerate}

\subsection{Converting bright time to dark time}

The polarising nature of the Lyot offers a very interesting prospect. In bright moonlight, 
the additional night sky from the moon is highly polarised ($>$80\%) at sight lines
perpendicular to the moon's direction. Therefore, the sky level in bands free of coronal 
emission can be greatly reduced by aligning the entrance polaroid of the Lyot with the 
moon's direction. For sky fields which lie at a lunar angle of 80-100\deg, the bright-moon 
sky could be reduced to dark-grey levels. Such an instrument could be used all year round 
to carry out cosmological surveys. For a detailed discussion of this principle, see
Baldry \& Bland-Hawthorn (2001).

\section{Conclusions}

To underscore the continuing importance of wide-field imaging surveys, CCD camera 
mosaics (typically 8K$\times$8K) are now operating or are under development at several 
major observatories. For the most part, these instruments are used for broadband imaging
although narrowband mosaic surveys have been performed (cf. Stiavelli\etal\ 2001;
Palunas\etal\ 2000).

It is very difficult to exploit a CCD mosaic spectroscopically in a fast beam. In
collaboration with K. Glazebrook, we have considered both objective grisms and 
transmissive volume-phase holographic gratings used in combination with doublet and 
triplet correctors. The transmissive grating produces a zeroth order which dramatically
increases the background over the field. The objective grism bypasses this problem but
at the expense of serious astigmatism across the field.

The tunable Lyot filter offers a powerful method for exploiting the full area of a 
large CCD mosaic. Traditionally, Lyot filters have been used in day-time astronomy, 
in atmospheric studies and in remote sensing. While half the light is lost at the first 
polarizing element, this is largely compensated for by the simpler optical system. 
The 5-stage, wide-field Lyot
proposed here has only 6 air-glass surfaces compared with $\sim$20 for the TAURUS-2
system (Taylor \& Atherton 1980). With recent improvements 
in producing birefringent elements, polaroids and wave plates,
the overall throughout will certainly compete with pre-existing imaging systems,
but with the advantage of much greater survey efficiency.

The construction and operation of a Lyot filter is moderately challenging. The primary
obstacles are the production of high quality achromatic polarizers, retarder elements, 
and wave plates, in order of increasing complexity.  The Lyot filter is tuned 
optomechanically which is a major advantage once the system is fully calibrated
(\eg temperature). For very large Lyot apertures ($>$100mm),  the wave plate problem 
may need to be handled in stages. The prototype could use simple retarder elements
which gives only about 50~nm of useful spectral coverage. Later devices could 
incorporate achromatic or superachromactic retarders. 

The Lyot filter is not the last word in wide-field tunable imaging.
\u{S}olc\footnote{The proper Czech pronunciation is `Sholtz'.} filters are 
highly non-intuitive tunable filters which use only two polarizers and a chain of 
identical retarders with varying position angles (\u{S}olc 1959; Evans 1958). 
There are many variants on the \u{S}olc principle, \eg the same can be achieved with 
polarizing filters by proper choice of crystal lengths.  There are folded (zigzag) 
and fanned designs with the former having the better performance (Beckers \&
Dunn 1965). Title (1999) has made a tunable \u{S}olc
filter with 70~mm clear aperture.  It has the extraordinary capability of
an adaptable spectral profile:  an $n$-element \u{S}olc filter can have a
profile that is determined by $n$ Fourier coefficients.  
To our knowledge, a wide-angle \u{S}olc filter using half-wave plates
has not been attempted although it is entirely feasible.

\acknowledgments
We would like to thank an anonymous referee for the suggested improvements to the
manuscript, many of which were deeply insightful.
JBH acknowledges a Long Term Consultancy at Lawrence Livermore National Labs (LLNL) 
where this work was started. We owe a particular debt to several people who were 
tempted out of retirement, including Harry Ramsey and W.H. `Beatty' Steel, to discuss 
aspects of this work.  JBH is indebted to Dick \& Alice Dunn, Ray Smartt, and the 
staff at Sunspot Observatory for their hospitality during the early stages of this 
work. We acknowledge crucial insights into the `lost art' of manufacturing bleached 
polaroids from Harry Ramsey, Lou Gilliam and Bruce Gillespie. JBH is grateful to Ian 
Thomas (LLNL) for discussions on crystal growth, and to Roger Netterfield, Chris 
Walsh, Zoltan Hegados and Bob Oreb at CSIRO (Lindfield) on the retardance and manufacture 
of crystal elements.  We would like to thank Alan Title, Bruce Woodgate and Jim Stilburn 
for their experimental insights, and Povilas Palunas and Gary da Costa for their generous 
assistance with several aspects of this work.  We acknowledge the kind assistance
of K. Shimasaku who calculated photometric redshifts, John Peacock and Shaun Cole for
useful discussions, and A. Diaferio who supplied us with a simulation from the GIF 
consortium.  The work by WvB was performed under the auspices of the U.S. Department of
Energy, National Nuclear Security Administration by the University of California,
Lawrence Livermore National Laboratory under contract No. W-7405-Eng-48.

\newpage
\begin{figure*}
\caption{
In a properly matched optical system, the ${\cal A}_1.\Omega_1$ product of the 
detector measured at the telescope aperture is equal to the ${\cal A}_2.\Omega_2$ 
product of the telescope measured at the detector. The telescope diameter is $D$ and 
the focal length of a spherical mirror is $f$; the detector size is $d$. The 
${\cal A}.\Omega$ invariant is the throughput or \'{e}tendue of the system.
\label{fig:A-Omega}}
\end{figure*}

\begin{figure*}
\caption{Angle convention for an off-axis ray incident on an optical element.
\label{fig:angles}}
\end{figure*}

\begin{figure*}
\caption[]{{\it Left:} A simulation of the on-axis response of a 
1\%$\lambda$ filter in different converging beams: f/1
(superfast), f/2 (Subaru Prime Focus), f/2.48 (UKST), f/3.3 (AAT
Prime Focus).  The telescope mirror is assumed to be a filled aperture.
For demonstration, we take the new UKST H$\alpha$
filter assuming a rectangular profile and $n_o=1.9$. Note the trends in
broadening (particularly FWZI), skewing, and the loss in mean transmission. 
The cross hairs show the predicted first and second moments from
equations~\ref{mean_delta} and \ref{quadrature}. {\it Right:} A
simulation of the response of the UKST H$\alpha$ 1\%$\lambda$ filter
in the f/2.48 beam, for the on-axis case (solid curve), and at an
off-axis position (thin curve) at the edge of the 14-inch filter.
\label{fig:on-axis}}
\end{figure*}

\begin{figure*}
\caption[]{The ray bundle incident on a filter in a converging beam
placed in front of a detector. In the on-axis case, the maximum
angle ($\theta_o$) is specified by the telescope beam. In the
off-axis case, the ray bundle is epicentric and depends both on 
the beam and the angular size of the detector ($\theta_f$). (The
contours are similar ellipses of low eccentricity in practice.)
\label{fig:rays}}
\end{figure*}

\begin{figure*}
\caption[]{A linearly polarized wave enters a birefringent element at normal incidence. 
In a positive uniaxial crystal, such as quartz, the optic axis is the fast axis. 
The $o-$ray has a faster speed through quartz than the $e-$ray (same $\nu$) and therefore a 
longer wavelength. The retardance between the rays builds up as the light propagates further
into the medium. The effect is grossly exaggerated here. Note that the light is elliptically 
polarized within the crystal and the plane of polarization changes slowly throughout the medium. 
The direction of the electric field is shown by $\vec{E}$.
\label{fig:biref}}
\end{figure*}

\begin{figure*}
\caption[]{A montage of Lyot stages demonstrating the different aspects of building
a Lyot filter. The elements are the polaroid (P), the retarder (R), the half-wave
plate ($\frac{\lambda}{2}$), and the quarter-wave plate ($\frac{\lambda}{4}$). The
black-tipped arrow shows the orientation of key elements; the white-tipped arrow 
indicates the direction of the fast axis; the optic axis runs horizontally. (i) The 
simplest Lyot filter producing the cosine behaviour in eq.~\ref{eq:xfr}. (ii) 
The wide-field compensated filter in which the retarder is split and crossed.
(iii) A single-stage filter with tuning capability. (iv) A wide-field compensated
filter with tuning capability. (v) A fragile, thin element filter with tuning 
capability. (vi) An equivalent unit to that presented in (v) with the aid of thicker 
elements through the differencing principle. The crossed retarders are sufficiently 
thick to require wide-field compensation.
\label{fig:montage}}
\end{figure*}

\begin{figure*}
\caption[]{The transmission profile of a simple Lyot cascade with (top to bottom)
1, 2, 3, 4 and 5 stages. The thinnest element has retardance R and the following
elements have thicknesses which are multiples of this element. The polaroids P are
aligned with each other and oriented at 45\deg\ to the retarders.
\label{fig:lyotprof}}
\end{figure*}

\begin{figure*}
\caption[]{A comparison of the Gaussian, Lorentzian and Lyot instrumental profiles
with matched full widths at half maxima. The 5-stage Lyot has much better wing
suppression than a Fabry-Perot device (e.g. TTF). The core of the Lyot profile is
well approximated by a Gaussian response function.
\label{fig:compprof}}
\end{figure*}

\begin{figure*}
\caption[]{The half-wave plate advances the phase of the $o$-ray by $\pi$ compared
to the $e$-ray. This is equivalent to a rotation of the electric field vector $\vec{E}$ by 
$\pi/2$.
\label{fig:halfwave}}
\end{figure*}

\begin{figure*}
\caption[]{The proximity of the Lyot filter to the Prime Focus in a telecentric system.
The hatched areas show the part of the filter which contributes to the image.
Lyots placed close to the detector can be smaller but this requires elements with high
image quality and uniformity. Lyots placed further away produce more even illumination 
over the detector although at the expense of much larger elements. {\it THE SHADED REGIONS ARE SUPPOSED TO BE
HATCHED.}
\label{fig:nonuniform}}
\end{figure*}

\begin{figure*}
\caption[]{The basic structure of the proposed 5-stage Lyot filter. The lowest order
stages serve as blocking filters to the higher orders. Stage I achieves a very thin
element through differenced, thick elements which have been crossed. Stages II and III 
are simple Lyot elements, where the thinner element is bonded to glass. Stages IV and V
use wide-field compensation. All stages are immersed in index-matched oil. Stage I and the 
outer polaroids are fixed; all other stages rotate differentially about the 
long axis. The upper rectangle shows the correct aspect ratio of the Lyot filter.
\label{fig:lowres}}
\end{figure*}

\begin{figure*}
\caption[]{Ray-traced designs for the Lyot 90mm and Gunn 6mm filters at the AAT f/3.3 prime focus used 
with conventional Wynne doublet and triplet correctors (top), and a new field-expanded doublet 
corrector (bottom). The details of the spot diagrams are summarised in Table~\ref{tab:spots}.
\label{fig:optics}}
\end{figure*}

\begin{figure*}
\vspace{10cm}
\caption[]{
Strip-mosaic scans of a subset of candidates from a {\sl TTF Field Galaxy 
Survey} field at $z\sim 0.1$.  Scan images ({\em top of each panel}) are 
9\arcsec\ on a side.
Circles show the integrating aperture used by the Sextractor photometry. 
TTF spectra ({\em bottom} of each panel) for the same galaxies are also shown, 
with initial ({\em dotted line}) and final ({\em solid line}) continuum fits. 
Numbers shown ({\em right}) are flux ($\times 10^{-16}$ erg cm$^{-2}$ s$^{-1}$
\AA$^{-1}$), star-galaxy classification parameter and $\sigma$-deviation.
Deviant points (excluded from the final continuum fit) are also
indicated ({\em circles}). The zero flux level is shown by the
horizontal tickmarks (where present) and non-detections are
represented on this level ({\em crosses}).  
\label{fig:collage}}
\end{figure*}

\begin{figure*}
\vspace{10cm}
\caption[]{
Unobscured H$\alpha$ ({\em left}) and [OII] line fluxes ({\em right}) for
galaxies with 0.1 -- 10 M$_\odot$\,yr$^{-1}$ as a function of redshift.
Star-formation rates for the Galaxy (5 M$_\odot$\,yr$^{-1}$; Smith, Biermann
\& Mezger 1978), the LMC (0.26 M$_\odot$\,yr$^{-1}$: Kennicutt\etal\ 1995),
and SMC (0.046 M$_\odot$\,yr$^{-1}$: Kennicutt\etal\ 1995) are shown for
comparison.
\label{fig:SFR}}
\end{figure*}

\begin{figure*}
\caption[]{A $\tau$CDM simulation by the GIF consortium (Kauffmann\etal\ 1999)
for which ($\Omega_m=0.8$, $\Omega_{\Lambda}=0.2$, $h=0.5$) 
normalised to the present cluster abundance ($\sigma_8 = 0.6$).
The simulation shows the expected distribution of galaxies with star formation
rates in excess of 0.3 M$_\odot$ yr$^{-1}$ in [OII] at $z\sim 1$.
The white points are the expected galaxies in a 2~nm band centered at 745~nm;
the black points are the additional objects in a band broadened to 6~nm.
The circle illustrates the TTF field size.
\label{fig:GIF}}
\end{figure*}

\begin{figure*}
\vspace{10cm}
\caption[]{
Galaxy colours at the redshifts of the major emission-lines within 
each TTF scan interval at 668, 707, 814 and 909~nm. The galaxy colour
tracks ({\em solid lines}) trace the range of galaxy types at the 
indicated redshift using the method of Fukugita\etal\ (1995).
The redshifts correspond to \Ha, \OIII, \Hb\ and 
\OII\ (in order of increasing redshift), for each spectral region.
The galaxy tracks close to each other are those for \OII\ and \Hb,
of which only \Hb\ is useful as a star-formation indicator.
\label{fig:phot-z}}
\end{figure*}

\newpage
\begin{table*}
\begin{tabular}{ccc}
$\cal F$ & $\lambda_N + \deltam \lambda_N$ & $\langle\sigma_\delta\rangle\,\lambda_N$ \\
& & \\ \hline
$\cal 1$ & 6590.0 & 19.0 \\
5.0 & 6585.4 &  19.2 \\
3.3 & 6579.5 &  20.0 \\
2.5 & 6571.6 &  21.8 \\
2.0 & 6561.2 &  25.2 \\
1.5 & 6538.5 &  35.1 \\
1.0 & 6471.8 &  69.7 \\
\end{tabular}
\caption{Predicted filter degradation for a 1\%$\lambda$ filter ($n_o = 1.9$)
in optical beams with a range of focal ratios.
\label{tab:degrade}}
\end{table*}

\clearpage
\begin{table*}
\begin{tabular}{ccccc}
Filter $\lambda/\Delta \lambda$ & \Ha & \OIII & \Hb & \OII \\
\hline
668/21 & 0.00   &  0.26 &  0.27 & 0.56 \\
707/26 & 0.08  	&  0.42	&  0.45	& 0.90 \\
814/33 & 0.24	&  0.63	&  0.67	& 1.18 \\
909/40 & 0.40 	&  0.82	&  0.87	& 1.44 \\
\end{tabular}
\caption{ Mean redshifts in the filters used for the {\it TTF Field Galaxy 
Survey}. The \OIII\ redshift is calculated from the mean wavelength of the 
$\lambda\lambda 495.9, 500.7$~nm doublet. 
\label{tab:lines}} 
\end{table*}

\clearpage
\begin{table*}
\begin{tabular}{rcccccccccc}
 & \multicolumn{5}{l}{\bf Lyot filter \rm (90mm thick)} &  
                       \multicolumn{5}{l}{\bf Gunn filter \rm (6mm thick)} \\
\\
$\lambda$ (nm)\ \ \ \ & 356 & 483 & 626 & 767 & 910\ \ \ \ & 336 & 437 & 580 & 716 & 864 \\
               &     &     &     &     &                   & 356 & 483 & 626 & 767 & 910 \\
               &     &     &     &     &                   & 376 & 529 & 672 & 818 & 956 \\
\\
 & \multicolumn{10}{c}{rms image diameters ($^{\prime\prime}$) } \\
\\
Field radius ($^\prime$)\ \ \ \  & \multicolumn{10}{l}{\bf Existing Wynne doublet} \\
\\
          0.0\ \ \ \  & 0.11 & 0.10 & 0.14 & 0.17 & 0.19\ \ \ \ & 0.35 & 0.21 & 0.14 & 0.10 & 0.08 \\
          8.8\ \ \ \  & 0.14 & 0.16 & 0.19 & 0.22 & 0.24\ \ \ \ & 0.32 & 0.20 & 0.14 & 0.11 & 0.10 \\
         12.5\ \ \ \  & 0.21 & 0.24 & 0.27 & 0.29 & 0.31\ \ \ \ & 0.27 & 0.19 & 0.15 & 0.15 & 0.16 \\
\\
 & \multicolumn{10}{l}{\bf Existing Wynne triplet} \\
\\
         0.0\ \ \ \  & 0.10 & 0.27 & 0.35 & {\underline{0.40}} & {\underline{0.42}}\ \ \ \ & 0.20 & 0.10 & 0.12 & 0.16 & 0.20 \\
        14.0\ \ \ \  & 0.10 & 0.34 & {\underline{0.47}} & {\underline{0.55}} & {\underline{0.59}}\ \ \ \ & 0.27 & 0.13 & 0.23 & 0.30 & 0.36 \\
        20.0\ \ \ \  & 0.13 & {\underline{0.40}} & {\underline{0.58}} & {\underline{0.68}} & {\underline{0.74}}\ \ \ \ & 0.34 & 0.16 & 0.31 & {\underline{0.41}} & {\underline{0.48}} \\
\\
 & \multicolumn{10}{l}{\bf New wide-field doublet} \\
\\
         0.0\ \ \ \  & 0.23 & 0.11 & 0.08 & 0.08 & 0.09\ \ \ \ & 0.36 & 0.26 & 0.21 & 0.20 & 0.20 \\
        14.0\ \ \ \  & 0.22 & 0.14 & 0.15 & 0.17 & 0.19\ \ \ \ & {\underline{0.52}} & {\underline{0.41}} & 0.26 & 0.22 & 0.21 \\
        20.0\ \ \ \  & 0.31 & 0.22 & 0.26 & 0.29 & 0.32\ \ \ \ & {\underline{0.58}} & {\underline{0.50}} & 0.35 & 0.35 & 0.37 \\
\end{tabular}
\caption{The performance of existing corrector elements with the Lyot filter and with 
{\it ugriz} filters. The performance of a new doublet design is also shown: this
produces good rms image sizes over a much wider field than the conventional Wynne doublet.
\label{tab:spots}}
\end{table*}
\clearpage

\appendix
\section{Narrowband filters in a converging beam}

From Snell's law and Abbe's principle,
\begin{equation}
 \left( \frac{\lambda_I}{\lambda_N} \right)^2 = 
  1 - \left( \frac{\sin \theta}{n_o} \right)^2
\end{equation}
where: $\lambda_I$ is the wavelength transmitted at an incidence 
angle of $\theta$ on the interference filter;
$\lambda_N$ is the wavelength transmitted at normal incidence, and
$n_o$ is the refractive index of the filter.
The fractional change in wavelength can be written as
\begin{eqnarray}
 \delta & = & \frac{\delta \lambda}{\lambda_N} \\ & = & \,
 -\! 1 + \frac{1}{n_o} \sqrt{n_o^2 - \sin^2 \theta} 
\end{eqnarray}
where $\delta \lambda = \lambda - \lambda_N$.
The change is always negative.

\smallskip
The centroid shift of a filter in a converging beam can be written as
\begin{eqnarray}
 \deltam & = & 
 \frac{\int \delta(\theta)\,I(\theta) \dd \theta}{\int I(\theta) \dd \theta} \\
 & = & 
 \frac{\int
 \left( -\! 1 + \frac{1}{n_o} \sqrt{n_o^2 - \sin^2 \theta} \right) 
 I(\theta) \dd \theta}{\int I(\theta) \dd \theta} 
 \label{eqn:centr-shift1}
\end{eqnarray}
where $I(\theta)$ is the intensity of light incident on the 
filter per unit angle.
For light focused from a reflecting mirror onto an on-axis filter,
$I(\theta) \propto \sin \theta$.
An appropriate substitution is, therefore, $\mu = \cos \theta$ 
and ${\rm d} \mu = - \sin \theta \dd \theta$\,.
Using this, Equation~\ref{eqn:centr-shift1} becomes
\begin{equation}
 \deltam \: = 
 \frac{\int -\! 1 + 
 \frac{1}{n_o} \sqrt{\mu^2 + n_o^2 - 1} \dd \mu}{\int {\rm d} \mu} \: .
 \label{eqn:centr-shift2}
\end{equation}

The limits for the integration come from the maximum and minimum ray angles. 
The maximum ray angle is related to the focal ratio ($\cal F$):
\begin{equation}
 \sin \theta_{\rm max} = \frac{1}{2 \cal F} 
\end{equation}
and the minimum ray angle $\theta_{\rm min} = 0$ if we ignore 
any central obstruction.
Equation~\ref{eqn:centr-shift2} with limits becomes
\begin{eqnarray}
 \deltam & = &
 \frac{1}{\mu_o - 1} \int_{1}^{\mu_o} -\! 1 + 
 \frac{1}{n} \sqrt{\mu^2 + m^2} \dd \mu \:  \\
 & = & 
 \frac{1}{\mu_o - 1} \left| -\! \mu + \frac{1}{2n_o} \mu \sqrt{\mu^2 + m^2} + 
 \frac{m^2}{2n_o} \ln \left( \mu + \sqrt{\mu^2 + m^2} \right)  \right|_1^{\mu_o}
\end{eqnarray}
where $\mu_o = \cos \theta_{\rm max}$ and $m^2 = n_o^2 - 1$, and
\begin{equation}
 \mu_o = \frac{1}{2 \cal F} \sqrt{4 {\cal F}^2 - 1} \: .
\end{equation}
Using $\sqrt{\mu_o^2 + m^2} = n_o(\delta_o + 1)$, we obtain
\begin{equation}
 \deltam \: = \frac{1}{2 (\mu_o - 1)} \left( 1 - \mu_o + \mu_o \delta_o +
 \frac{m^2}{n_o} \ln \left[ \frac{\mu_o +  n_o\delta_o + n_o }{n_o + 1} \right] \right)
\end{equation}
where $\delta_o$ is the fractional change in wavelength for the 
maximum ray angle, \ie
\begin{equation}
 \delta_o = -\!1 + \frac{1}{n_o} \sqrt{\mu_o^2 + m^2} \: .
\end{equation}
Alternatively, 
\begin{equation}
 \deltam \: = 
 \half \left( -\! 1 + \delta_o + \frac{\delta_o}{\delta_o'} \right)
 + \frac{m^2}{2 n_o \delta_o'} 
 \ln \left( 1 + \frac{\delta_o' + n_o\delta_o}{n_o+1} \right) \: .
\end{equation}
where $\delta_o' = \mu_o - 1$.  With $n_o=1$, 
\begin{eqnarray}
\deltam & = & \frac{\delta_o}{2} \\ & = & \frac{\mu_o - 1}{2} \: .
\end{eqnarray}

The variance can also be calculated:
\begin{eqnarray}
\sigma_{\delta}^2 & = & \frac{\int \delta(\theta)^2\,I(\theta) \dd \theta}
 {\int I(\theta) \dd \theta} - \deltam^2 
\\
& = & \frac{1}{\mu_o - 1} \int_{1}^{\mu_o} \left( -\! 1 + 
 \frac{1}{n_o} \sqrt{\mu^2 + m^2} \right)^2 \dd \mu \: - \deltam^2
\\
& = &
 \frac{(\mu_o  + 2)(\mu_o - 1)}{3n_o^2}
 \: - 2 \deltam - \deltam^2 \: .
\end{eqnarray}

With $n_o=1$,
\begin{eqnarray}
 \sigma_{\delta}^2 & = & \frac{\delta_o^2}{12} \\ 
& = & \frac{(\mu_o - 1)^2}{12} \: .
\end{eqnarray}

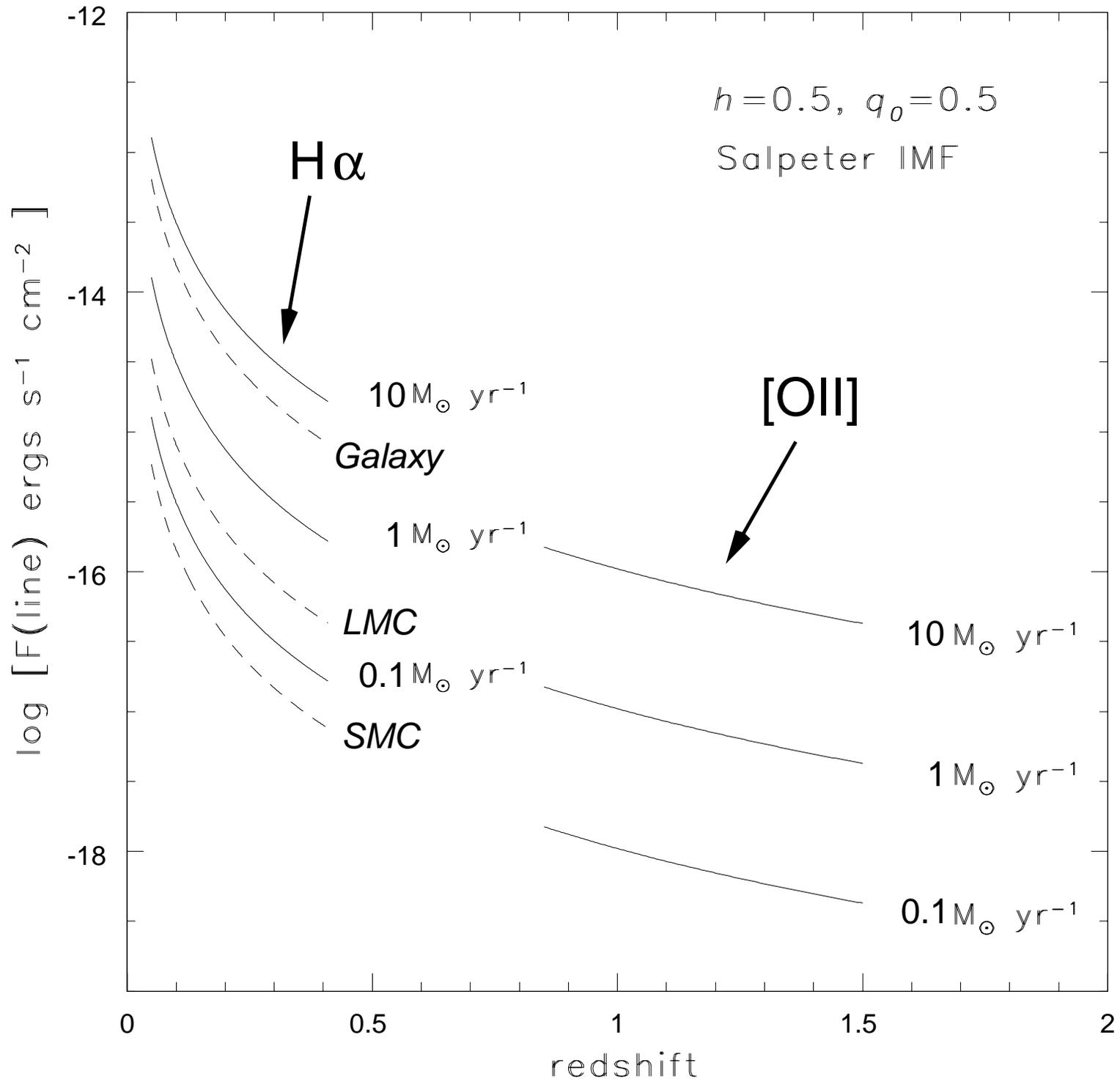
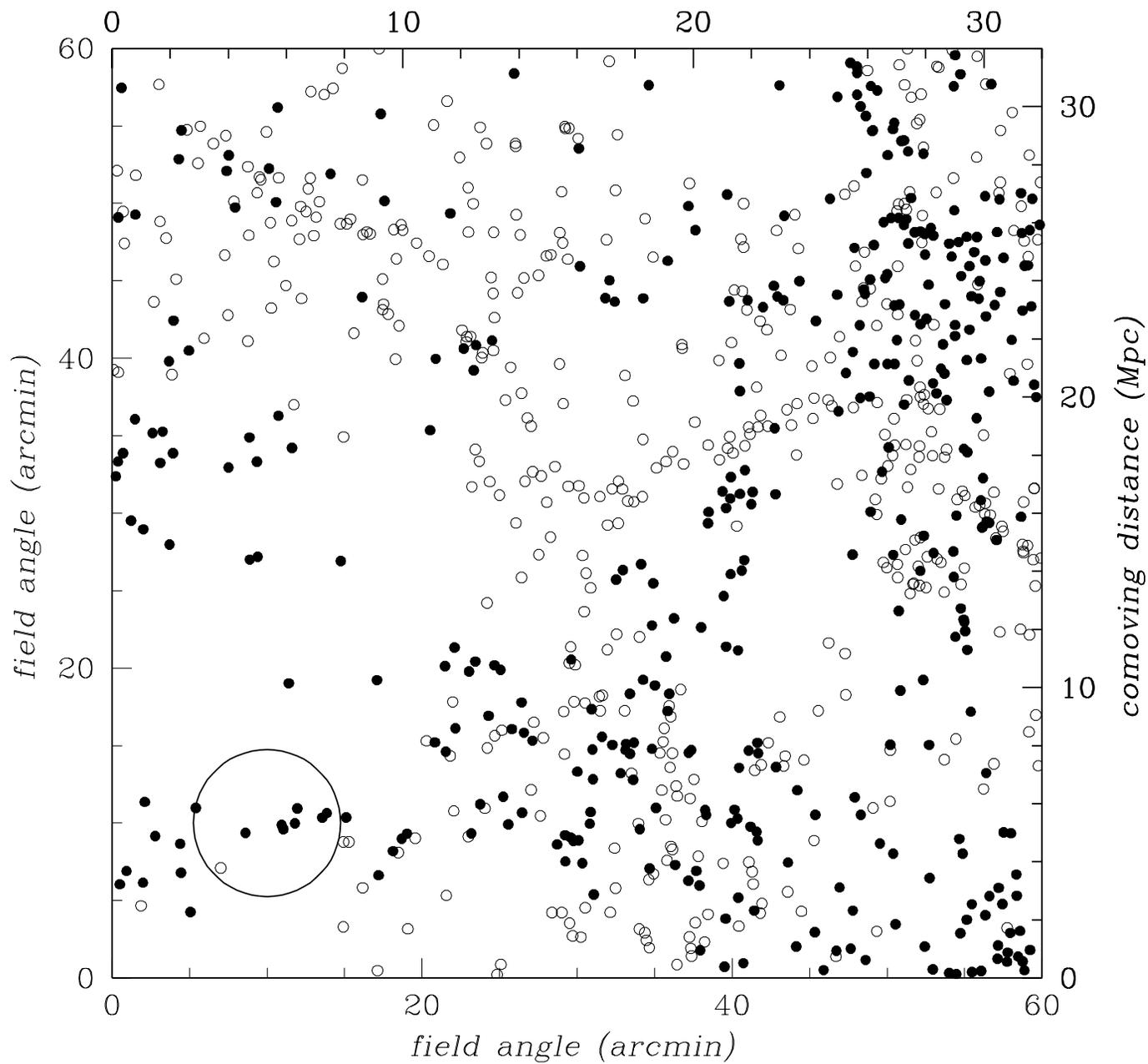
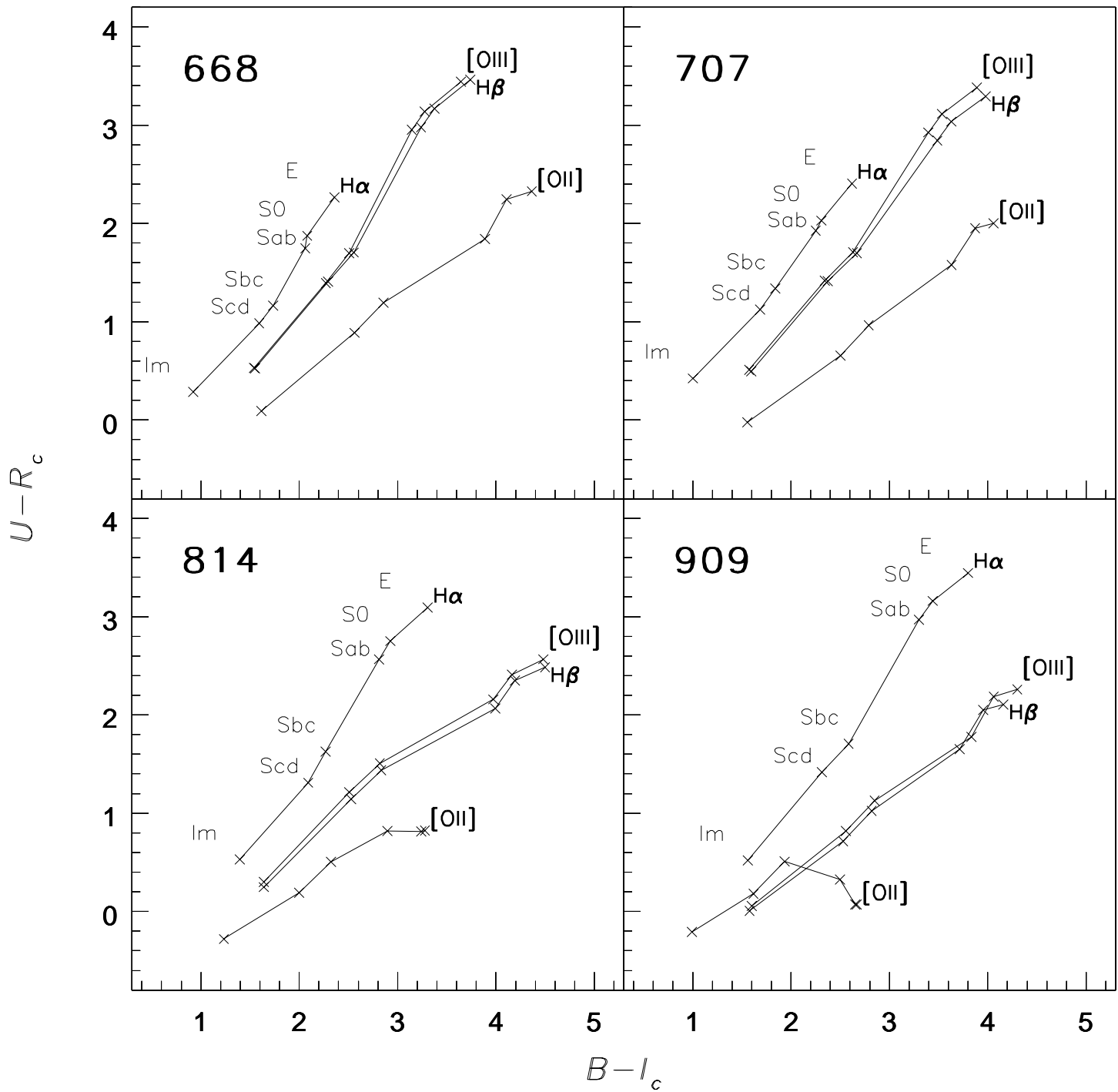
\end{document}